\documentclass{rspublic}

\usepackage{graphicx}
\usepackage{gastex}

\newcommand{\beq}[1]{  \\ {\tiny ({#1})}   \begin{equation} \label{#1} }
\newcommand{\beqa}[1]{ \\ {\tiny ({#1})}   \begin{eqnarray} \label{#1} }
\renewcommand{\beq}[1]{\begin{equation} \label{#1} }\renewcommand{\beqa}[1]{\begin{eqnarray} \label{#1} }
\newcommand{\eeq}{\end{equation}}
\newcommand{\eeqa}{\end{eqnarray}}
\newcommand{\ba}[1]{\begin{array}{#1}}		\newcommand{\ea}{\end{array}}
\newcommand{\rf}[1]{(\ref{#1})}

\begin{document}

\title[Poisson's ratio in cubic materials]{Poisson's ratio in cubic materials}

\author[A. N. Norris]{Andrew N. Norris}

\affiliation{Mechanical and Aerospace Engineering, 
	Rutgers University, Piscataway NJ 08854-8058, USA \,\, norris@rutgers.edu}

\label{firstpage}

\maketitle

\begin{abstract}{Poisson's ratio, cubic symmetry, anisotropy}

Expressions are given for the maximum and minimum  values of Poisson's ratio $\nu$ for materials with cubic symmetry.  Values less than $-1$  occur if  and only if the maximum shear modulus is associated with the cube axis and is at least $25$ times  the value of the minimum shear modulus.   Large values of $|\nu|$ occur in directions at which the Young's modulus is approximately equal to one half of its $111$ value.   Such directions, by their nature, are very close to $111$.  Application to data for cubic crystals indicates that certain Indium Thallium alloys simultaneously exhibit Poisson's ratio less than -1 and greater than +2.

\end{abstract}

\section{Introduction}

The Poisson's ratio  $\nu$ is an important physical quantity in the mechanics of solids, second only in significance to the Young's modulus.  It is strictly bounded  between  $-1$ and $1/2$ in isotropic solids, but no such simple bounds exist for anisotropic solids, even for those closest to isotropy in material symmetry: cubic materials.  In fact, 
Ting \& Chen (2005) demonstrated that arbitrarily large positive and negative values of Poisson's ratio could occur in solids with cubic material symmetry.  The key requirement is that the Young's modulus in the $111-$direction is  very large (relative to other directions), and as a consequence the Poisson's ratio for stretch close to but not coincident with the $111-$direction can be large, positive or negative.    Ting \& Chen's   result replaces  conventional wisdom (eg. Baughman {\it et al.} 1998) that the extreme values of  $\nu$ are associated with stretch along the face diagonal ($110-$direction).  Boulanger \& Hayes (1998) showed that  arbitrarily large values of  $|\nu|$ are possible in materials of orthorhombic symmetry.  
Both pairs of authors analytically constructed  sets of elastic moduli which show the unusual properties while still physically admissible. 
The dependence of the large values of Poisson's ratio on elastic moduli and the related scalings of strain are discussed by Ting (2004) for cubic and more  anisotropic materials. 

To date there is no  anisotropic elastic symmetry  for which there are analytic expressions of the extreme values of  Poisson's ratio for all materials in the symmetry class, although  bounds may be obtained for some specific pairs of directions for certain material symmetries.  For instance, 
Lempriere (1968) considered Poison's ratios for stretch and transverse strain along the principal directions, and showed that the  is bounded by the square root of the ratio of principal Young's moduli, $|\nu ({\bf n},{\bf m})| < (E({\bf n})/E({\bf m}))^{1/2}$ (in the notation defined below).  Gunton and Saunders (1975) performed some numerical searches for the extreme values of $\nu$ in materials of cubic symmetry.  However, the larger question of  what limits on $\nu$ exist for all possible pairs of directions remains open, in general. 
This paper  provides an answer for materials of cubic symmetry.  Explicit formulas are obtained for the minimum and maximum values of $\nu$ which allow us to examine  the occurrence of the unusually large values of Poisson's ratio, and the conditions under which they appear.  Conversely, we can also define the range of material parameters for which the extreme values are of ``standard" form, i.e. associated with principal pairs of directions  such as 
$\nu (110,1\bar{1}0)$ for stretch and measurement along the two face diagonals. For instance, we will see that a necessary condition that  one or more of the  extreme values of Poisson's ratio  is not associated with a principal direction is that $\nu (110,1\bar{1}0)$ must be less than $-1/2$.   The general results are also illustrated by application to a wide variety of cubic materials, and it will be shown that values of $\nu < -1$ and $\nu >2$ are possible for certain stretch directions in existing solids. 

We begin in \S\ref{sec2} with definitions of moduli and  some preliminary results.  
An important identity is presented which enables us to obtain the extreme values of both the shear modulus  and Poisson's ratio  for a given choice of the extensional direction.  Section \ref{sec4}
considers the central problem of obtaining extreme values of $\nu$ for all possible pairs of orthogonal directions.  The solution  requires several new quantities, such as    
the values of $\nu$ associated with principal direction pairs. Section \ref{sec5} describes the range of possible elastic parameters consistent with positive definite strain energy.  The explicit formulae the global extrema are presented 
 and their overall properties are discussed in \S\ref{sec6}.  It is shown that certain Indium Thallium alloys simultaneously display values of $\nu$ below $-1$ and above $+2$.

\section{Definitions and preliminary results}\label{sec2}

The fourth order tensors of compliance and stiffness   for a cubic material,  ${\bf S} $ and ${\bf C}= {\bf S}^{-1}$, may be written (Walpole 1984) in terms of three moduli $\kappa$, $\mu_1$ and $\mu_2$, 
\beq{sc}
{\bf S}^{\pm 1}  =  (3\kappa)^{\mp 1} \, {\bf J} + (2\mu_1)^{\mp 1} \, \big( {\bf I}- {\bf D}\big)
+(2\mu_2)^{\mp 1} \, \big( {\bf D}- {\bf J}\big).
\eeq
Here $I_{ijkl} = \frac12 (\delta_{ik}\delta_{jl}+ \delta_{il}\delta_{jk})$ is the fourth order identity, 
$J_{ijkl} = \frac13 \delta_{ij}\delta_{kl}$, and 
\beq{D}
D_{ijkl} = \delta_{i1}\delta_{j1} \delta_{k1}\delta_{l1} +
\delta_{i2}\delta_{j2} \delta_{k2}\delta_{l2}+
\delta_{i3}\delta_{j3} \delta_{k3}\delta_{l3}\, .
\eeq
The isotropic tensor ${\bf J}$ and the tensors of cubic symmetry $\big( {\bf I}- {\bf D}\big)$ and $\big( {\bf D}- {\bf J}\big)$ are  positive definite (Walpole 1984), so  the requirement of positive strain energy is that $\kappa$, $\mu_1$ and $\mu_2$ are positive.  These three parameters, called the ``principal elasticities" by Kelvin (Thomson 1856),  can be related to  the  standard Voigt  stiffness  notation: $
\kappa = (c_{11}+2c_{12})/3$, $\mu_1 = c_{44}$ and $\mu_2 = (c_{11}-c_{12})/2$.  Alternatively, 
$
\kappa = (s_{11}+2s_{12})^{-1}/3$, $\mu_1 = s_{44}^{-1}$ and $\mu_2 = (s_{11}-s_{12})^{-1}/2$ in terms of the compliance. 


Vectors, which are usually unit vectors, are denoted by lowercase boldface, e.g. $\bf n$.  
The triad $\{ {\bf n}, {\bf m}, {\bf t} \}$ represents an arbitrary  orthonormal  set of vectors.  
Directions are  also  described using crystallographic notation, e.g. ${\bf n}=1\bar{1}0$ is the unit vector $(1/\sqrt{2},-1/\sqrt{2},0)$.  The summation convention on repeated indices is assumed.

\subsection{Engineering moduli}

The {\it Young's modulus} $ E( {\bf n} )$ sometimes written $E_{\bf n}$,  {\it shear modulus} $ G( {\bf n}, {\bf m} )$  and {\it Poisson's ratio} $ \nu( {\bf n}, {\bf m} )$ are    
(Hayes 1972) 
\beq{e4}
 E( {\bf n} ) = {1}/{s_{11}'} , \qquad G( {\bf n}, {\bf m} )= {1}/{s_{44}'}
 , \qquad \nu ( {\bf n}, {\bf m} ) = -{s_{12}'}/{s_{11}'}
 \, , 
 \eeq
 where $s_{11}' = s_{ijkl}n_in_jn_kn_l$, $s_{12}' = s_{ijkl}n_in_jm_km_l$ and    
$s_{44}' = 4s_{ijkl}n_im_jn_km_l$.  Thus, $E( {\bf n} )$ and $\nu ( {\bf n}, {\bf m} )$ are defined by the axial and orthogonal strains in the $\bf n$ and $\bf m$ directions, respectively,  for an applied uniaxial stress in the $\bf n$ direction. $E$ and $G$ are positive  while $\nu$ can be of either sign or zero. A material for which $ \nu <0 $ is called auxetic, a term apparently introduced by K. Evans in 1991.  Gunton and Saunders (1975) provide an earlier but informative historical perspective on Poisson's ratio.   Love  (1944) reported a 
  Poisson's ratio of ``nearly $-1/7$" in Pyrite, a cubic crystalline material.

The  tensors ${\bf I}$ and ${\bf J}$ are isotropic, and consequently the directional dependence of the engineering quantities is through ${\bf D}$.  Thus, 
\beqa{egnu}
\frac{1}{E} &=& \frac{1}{9\kappa}+ \frac{1}{3\mu_2} - \big( \frac{1}{\mu_2} - \frac{1}{\mu_1} \big)
\, F({\bf n}), 
\\
\frac{1}{G} &=& \frac{1}{\mu_1} +\big( \frac{1}{\mu_2} - \frac{1}{\mu_1} \big)
\, 2D({\bf n}, {\bf m}),
\label{egnub}
\\
\frac{\nu}{E} &=& -\frac{1}{9\kappa}+ \frac{1}{6\mu_2} - \big( \frac{1}{\mu_2} - \frac{1}{\mu_1} \big)
\, \frac12 D({\bf n}, {\bf m}), \label{egnuc}
\eeqa
where 
\beq{df}
F ({\bf n}) =  n_1^2n_2^2 + n_2^2n_3^2 + n_3^2n_1^2 , 
\qquad 
D ( {\bf n}, {\bf m} ) = n_1^2m_1^2 + n_2^2m_2^2 + n_3^2m_3^2   
 \, . 
\eeq
We note for future reference the relations 
\beq{fol}
D ( {\bf n}, {\bf m} ) + D ( {\bf n}, {\bf t} )
 = 1 - (n_1^4+n_2^4+n_3^4)
 = 2F ({\bf n}) \, . 
\eeq

\subsection{General properties of $E$, $G$ and related moduli}

Although  interested primarily in the  Poisson's ratio, we first discuss some general results for $E$, $G$ and related quantities in cubic materials: the  area modulus $A$, and  the traction-associated bulk modulus $K$, defined below. 
The extreme values of $E$ and $G$ follow from the fact that 
$0\le F \le 1/3$ and $0\le D \le 1/2$  (Walpole 1986; Hayes \& Shuvalov 1998).  Thus, 
$G_{\rm min,\, max} $ $= {\rm min,\, max}(\mu_1,\, \mu_2)$, 
$E_{\rm min,\, max}$  $= 3\, \big[ ({3\kappa})^{-1} + {G_{\rm min,\, max}^{-1} }\big]^{-1}$, 
and  $E_{\rm min},E_{\rm max} = E_{001},E_{111}$ for $\mu_1 >\mu_2$, with the values reversed for $\mu_1 <\mu_2$ (Hayes \& Shuvalov 1998). 
As noted by Hayes \& Shuvalov (1998), the difference in extreme values of $E$ and $G$ are related by 
\beq{un}
3/E_{\rm min}- 3/E_{\rm max} = 1/G_{\rm min}  - 1/G_{\rm max} \,. 
\eeq
The extreme values also  satisfy 
\beq{099}
3/E_{\rm min,max}-1/G_{\rm min,max} = 1/ ({3\kappa})\, . 
\eeq
The shear modulus $G$ achieves both minimum and maximum values if ${\bf n}$ is directed along face diagonals, that is, $G_{\rm min}\le G \le G_{\rm  max}$ for ${\bf n}=110$. 

The {\it area modulus of elasticity} $A({\bf n})$ for the plane orthogonal to $\bf n$ is the ratio of an equibiaxial stress to the relative area change in the plane in which the stress acts (Scott 2000). Thus,  $1/A({\bf n}) = s_{ijkl}(\delta_{ij} - n_in_j)(\delta_{kl} - n_kn_l)$.  Using the equations above it may be shown that, for a cubic material,  
\beq{A}
1/{A({\bf n})} -1/{E({\bf n})}  =  1/(3\kappa)\, . 
\eeq
The   {\it averaged Poisson's ratio} $\overline{\nu}({\bf n})$ is defined as the average over $\bf m$ in the orthogonal  plane, or $\overline{\nu}({\bf n}) =  [\nu ({\bf n}, {\bf m})+\nu ({\bf n}, {\bf t})]/2$. 
The following result, apparently first obtained by 
Sirotin \&  Shaskol'skaya (1982), follows from the relations \rf{fol}, 
\beq{e8}
\big[{1 - 2 \overline{\nu}({\bf n})}\big]/{ E( {\bf n} ) }  = 1/(3\kappa )\, .   
\eeq
Equation \rf{e8} indicates that the extrema of $\overline{\nu}({\bf n})$ and $E({\bf n})$ coincide.  
The {\it traction-associated bulk modulus} $K({\bf n})$, introduced by He (2004),  relates the uniaxial stress in the $\bf n$ direction to the relative change in volume in  anisotropic materials. It  is defined by $3K({\bf n}) = 1/s_{iikl}n_kn_l$, and for cubic materials is simply
$K({\bf n}) = \kappa$. 
It is interesting to note that the relations \rf{099} through \rf{e8} have the same form as for  isotropic materials, for which $E$, $G$, $\nu$, $A$ and $K$ are constants. 
Equations \rf{egnu} to \rf{egnuc} imply other  identities, e.g. that 
the combination $1/{G } + 4 \nu /{E} $ is constant.  

Further discussion of  the extremal properties of $G$ and $\nu$ requires knowledge of how they vary with $\bf m$ for given $\bf n$, and in particular, the extreme values as a function of $\bf m$ for arbitrary $\bf n$, considered in the next subsection.  Note that   ${\bf n} =111$ and ${\bf n} = 001$ are the only directions for which $\nu ( {\bf n}, {\bf m})$ and $G ( {\bf n}, {\bf m})$ are independent of $\bf m$.  
It will become evident that ${\bf n} = 111$ is a critical direction, and we therefore 
 rewrite $E$ and $\nu$ in forms  emphasizing this direction:
\beq{sug}
\frac{1}{E({\bf n})} = \frac{1}{E_{111}} +   \big[ \frac13 - F({\bf n}) \big]\, \chi \, , 
\qquad 
\frac{\nu({\bf n}, {\bf m})}{E({\bf n})} =  \frac{\nu_{111}}{E_{111}} +  \big[\frac13 -  D({\bf n}, {\bf m}) \big]\,\frac{\chi}{2} \,  ,
\eeq
where $E_{111} = E(111)$, $\nu_{111} = \nu(111,\cdot)$ and 
$\chi$ (Hayes \& Shuvalov 1998) are
\beq{111}
E_{111} = \bigr(\frac{1}{9\kappa} +  \frac{1}{3\mu_1}\bigr)^{-1},
\qquad
\nu_{111} = \frac{3\kappa-2\mu_1}{6\kappa+2\mu_1},
\qquad \chi = \frac{1}{\mu_2} -  \frac{1}{\mu_1}. 
\eeq
Both $E_{111}$ and $\nu_{111}$ are independent of $\mu_2$. The fact that $F \le 1/3$ with equality  for  ${\bf n} = 111$ implies that this is  is the only stretch direction for which $E$, and hence $\nu$, are independent of $\mu_2$.   Equations \rf{sug} indicate that  $E({\bf n})$ and $\nu({\bf n}, {\bf m})$  depend  on  $\mu_2$ at any point in the neighbourhood of $111$, with  particularly strong dependence if  $\mu_2$ is small.   This singular behaviour is the reason for the extraordinary values of $\nu$ discovered by Ting \&  Chen (2005) and will be discussed at  further length below after we have determined the global extrema for $\nu$. 

\subsection{Extreme values of $G$ and $\nu$ for fixed $\bf n$}\label{sec3}

For a given  $\bf n$ consider the defined  vector
\beq{mm}
{\bf m}(\lambda) \equiv \rho \, \big(
\frac{n_1}{n_1^2 - \lambda},\, \frac{n_2}{n_2^2 - \lambda},\,\frac{n_3}{n_3^2 - \lambda} \big),
\eeq
with    $ \rho$ chosen to make ${\bf m}$ a unit vector.
Requiring  ${\bf n}\cdot {\bf m} = 0 $  implies that ${\bf m}(\lambda)$ is orthogonal to $\bf n$ if 
\beq{101}
\frac{n_1^2}{n_1^2 - \lambda}+\frac{n_2^2}{n_2^2 - \lambda}+\frac{n_3^2}{n_3^2 - \lambda}=0,
\eeq
{i.e.}, if $\lambda$ is a root of the quadratic
\beq{mm3}
\lambda^2 - 2\lambda (n_1^2n_2^2 + n_2^2n_3^2 + n_3^2n_1^2) + 3 n_1^2n_2^2n_3^2 
= 0.
\eeq
It is shown in Appendix A that the extreme values of $D ( {\bf n}, {\bf m} )$ for fixed $\bf n$ coincide with these roots, which are non-negative, and that the corresponding unit $\bf m$ vectors provide the extremal lateral directions. The basic result is described next.

\subsubsection*{A fundamental result:}
 Let   $0\le \lambda_-\le \lambda_+\le 1/2$ be the roots of \rf{mm3} and ${\bf m}_-,\, {\bf m}_+$ the associated vectors from \rf{mm}, i.e.  
 \begin{subequations}\label{sum}
\begin{align}\label{mm4}
\lambda_\pm  =&  (n_1^2n_2^2 + n_2^2n_3^2 + n_3^2n_1^2) \pm
\sqrt{ (n_1^2n_2^2 + n_2^2n_3^2 + n_3^2n_1^2)^2- 3 n_1^2n_2^2n_3^2 },
\\ 
{\bf m}_\pm = & \rho_\pm \, \big(
\frac{n_1}{n_1^2 - \lambda_\pm },\, \frac{n_2}{n_2^2 - \lambda_\pm },\,\frac{n_3}{n_3^2 - \lambda_\pm } \big),\, 
\\
\rho_\pm =&  \big[
\frac{n_1^2}{(n_1^2 - \lambda_\pm)^2} +\frac{n_2^2}{(n_2^2 - \lambda_\pm)^2}+\frac{n_3^2}{(n_3^2 - \lambda_\pm)^2}
 \big]^{-1/2} .  
\end{align}
\end{subequations}
The extreme values of $D$ for given  ${\bf n}$ are $\lambda_\pm $ associated with the orthonormal triad $\{ {\bf n}, {\bf m}_-, {\bf m}_+\}$, i.e.
\beq{dmm}
D_{\rm min} ( {\bf n}) = D ( {\bf n}, {\bf m}_- ) = \lambda_-, \qquad  D_{\rm max}( {\bf n}) = D ( {\bf n}, {\bf m}_+ )=\lambda_+ .
\eeq 
The extreme values of $G$ and $\nu$ for fixed $\bf n$ follow from eqs. \rf{egnub} and \rf{egnuc}.

\begin{figure}[ht]
				\begin{center}	
				\includegraphics[width=4.0in , height=2.8in 					]{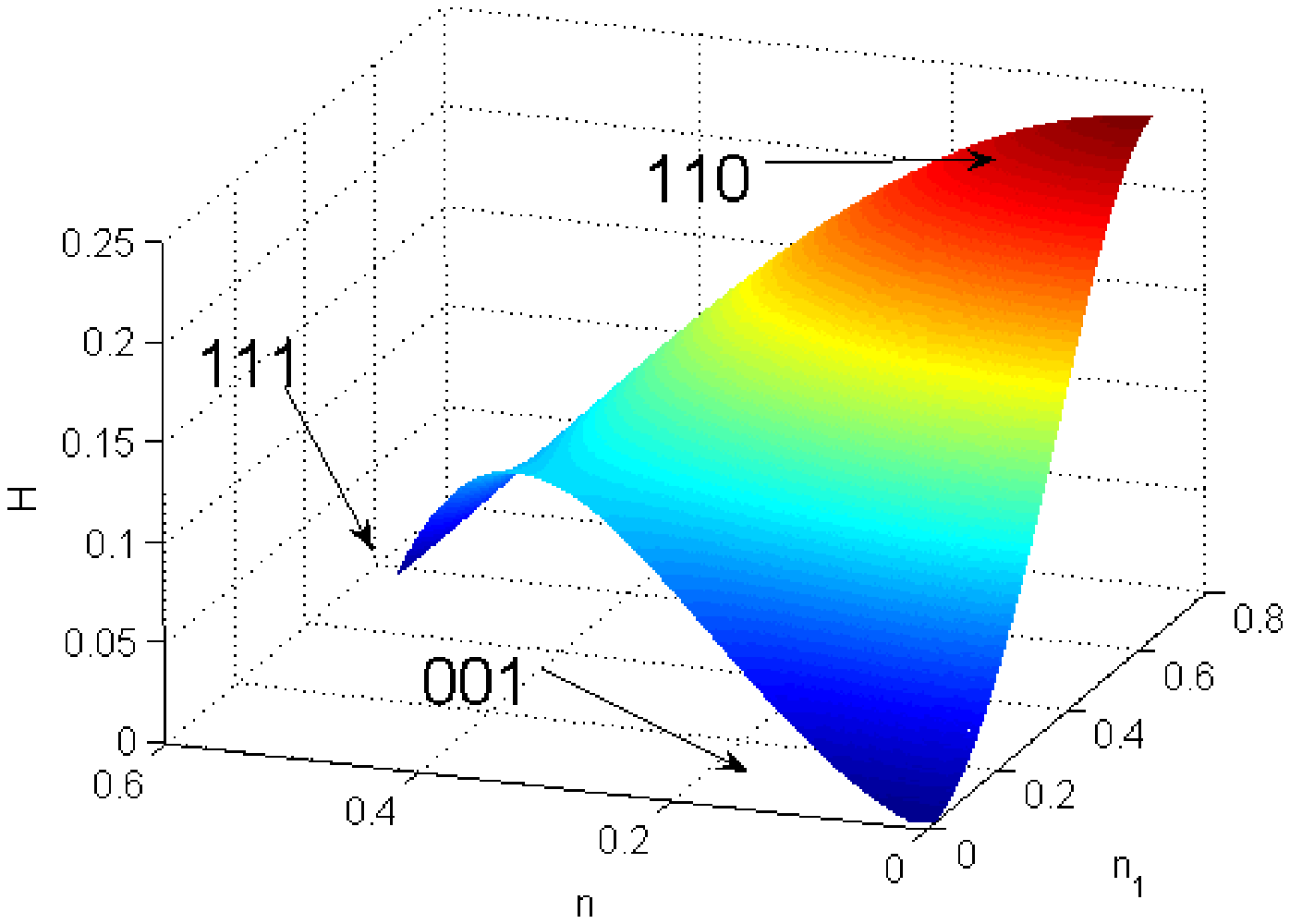} 
	\caption{The function $H$ of equation \rf{r2a} plotted vs. $n_1$ and $n_2$ for the region of solid angle depicted in figure \ref{fcube}.   Vertices ${\bf n} = 111$, $110$ and $001$ are indicated.	$H$ vanishes at $111$ and $001$ and is positive elsewhere, with maximum of $1/4$ along ${\bf n} = 110$ (face diagonals).  }
		\label{plotH} \end{center}  
	\end{figure}

 
 The above result also implies that the 
 extent of the variation  of the shear modulus and the Poisson's ratio for a given stretch direction $\bf n$ are  
 \begin{subequations}
 \begin{align}\label{q1}
1/{G_{\rm min}({\bf n}) } - 1/{G_{\rm max}({\bf n}) } &= \left|
 \chi \right| \, 4H({\bf n}), 
 \\ 
  \nu_{\rm max}({\bf n}) -  \nu_{\rm min } ({\bf n})&= \left| \chi \right| \, E({\bf n})H({\bf n})
 \, ,   \label{q1b}
\end{align}
\end{subequations}
where $H({\bf n})$ is, see figure \ref{plotH},  
\beq{r2a}
H ({\bf n}) = \big[ (n_1^2n_2^2 + n_2^2n_3^2 + n_3^2n_1^2)^2 - 3n_1^2n_2^2n_3^2\big]^{1/2} \, .  
\eeq 

\section{Poisson's ratio}\label{sec4}

\begin{figure}[htbp]
				\begin{center}	
 	
\setlength{\unitlength}{.2in}
\begin{picture}(10,10)(0,0)
\drawline[AHnb=0,linewidth=.05](0,0)(8,0)(8,8.764)(0,8.764)(0,0)
\drawline[AHnb=0,linewidth=.05](0,8.764)(4,10.764)(12,10.764)(12,2)(8,0)
\drawline[AHnb=0,linewidth=.05](8,8.764)(12,10.764)
\drawline[AHnb=0,linewidth=.1](4,4.382)(8,8.764)(8,4.382)(4,4.382)
\drawline[AHnb=1,AHLength=.5,linewidth=.1](8,8.764)(10,9.764)
\put(6.1,3.6){\makebox(0,0){\bf{\large{1}}}}
\put(5.5,6.7){\makebox(0,0){\bf{\large{2}}}}
\put(7.7,5.8){\makebox(0,0){\bf{\large{3}}}}
\put(9.2,8.6){\makebox(0,0){\bf{\large{3$^\prime$}}}}
\put(3.1,3.8){\makebox(0,0){[100]}}
\put(9.0,3.8){\makebox(0,0){[110]}}
\put(7.3,9.2){\makebox(0,0){[111]}}
\end{picture}
	\caption{The irreducible $1/48 $th of the cube surface is defined by the isosceles triangle with edges 1, 2 and 3.  The vertices opposite these edges correspond to, ${\bf n} = 111$, $110$ and $001$, respectively. 	Note that the edge 3' is equivalent  to 3 (which is used in  Appendix B). }
		\label{fcube} \end{center}  
	\end{figure}
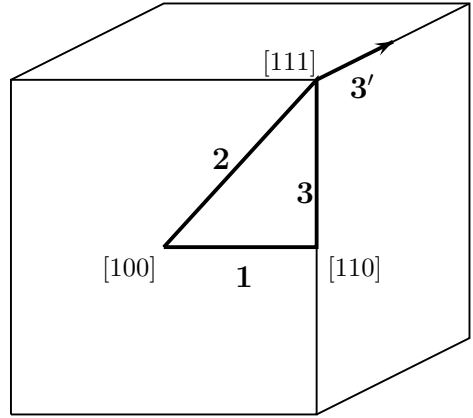

We now consider the global extrema of $ \nu ( {\bf n}, {\bf m} )$ over all directions $ 
 {\bf n}$ and ${\bf m}$.   Two methods are used to derive the main results.  The first 
uses general equations for a stationary value of $\nu$ in anisotropic media to  obtain a single equation which  must be satisfied if the stationary value lies in the interior of the triangle in figure \ref{fcube}.  It is shown that this condition, which  is independent of material parameters, is not satisfied, and hence    all  stationary values of $\nu$ in cubic materials  lie on the edges of the triangle.  This simplifies the problem considerably, and  permits us to deduce explicit   relations for the stationary values.   The second method, described in Appendix B, confirms the first approach by a comprehensive numerical test of all possible material parameters.  
 
\subsection{General conditions for stationary Poisson's ratio}

General conditions can be derived which must be satisfied in order that Poisson's ratio is stationary in  anisotropic elastic materials (Norris 2006).   These are: 
\beq{3cond}
s_{14}'= 0, 
\qquad
2\nu s_{15}' + s_{25}'= 0,
\qquad
(2\nu -1 )s_{16}' + s_{26}'= 0, 
\eeq
where the stretch is in the $1'$ direction $({\bf n})$ and $2'$ is the lateral direction $({\bf m})$. 
The conditions may be obtained by considering the derivative of $\nu$ with respect to rotation of the pair $( {\bf n},{\bf m})$ about an arbitrary axis.  Setting the derivatives to zero yields the stationary conditions \rf{3cond}. 

The only non-zero contributions to $s_{14}'$, $s_{15}'$, $s_{25}'$,  $s_{16}'$ and $s_{26}'$ in a  material of cubic symmetry come from $\bf D$.   Thus, we may rewrite the conditions for stationary values of $\nu$ in terms of $D_{14}' = D_{1123}'$ etc., as
\beq{cond}
D_{14}' = 0, 
\qquad
2\nu D_{15}' + D_{25}' = 0, 
\qquad
(2\nu -1 )D_{16}' + D_{26}' = 0. \
\eeq
The first  is automatically satisfied by virtue of the choice of the direction ${\bf m}$ as either of ${\bf m}_\pm $.  Regardless of which is chosen,  
\begin{align}\label{3c1}
D_{14}' &= D_{ijkl}n_in_jm_{+k}m_{-l}
\nonumber \\ &= 
\rho_+ \rho_- \bigg[ 
\frac{n_1^4}{(n_1^2 - \lambda_+)(n_1^2 - \lambda_-)} +
\frac{n_2^4}{(n_2^2 - \lambda_+)(n_2^2 - \lambda_-)} +
\frac{n_3^4}{(n_3^2 - \lambda_+)(n_3^2 - \lambda_-)} 
\bigg] 
\nonumber \\ &=   0 \, . 
\end{align}
The final identity may be derived  by first splitting each term into partial fractions and using the following (see Appendix A)
\beq{3c2}
\frac{n_1^4}{n_1^2 - \lambda_\pm} +
\frac{n_2^4}{n_2^2 - \lambda_\pm} +
\frac{n_3^4}{n_3^2 - \lambda_\pm} 
 = 1\, .
 \eeq
 
With no loss in generality, consider the specific case of ${\bf m} = {\bf m}_a$, $\lambda = \lambda_a$ where $a=+$ or $a=-$, and in either case,  $b=-a$.  It may be shown without much difficulty (Appendix B) that $\rho_\pm >0$ for $\bf n$ in the interior of the triangle of  figure \ref{fcube}.  It then follows that inside the triangle, 
\beq{490}
\frac{D_{15}'}{\rho_b} = \frac{D_{16}'}{\rho_a} = 1,
\quad
\frac{D_{25}'}{\rho_b} =\frac{\lambda_a}{\lambda_a-\lambda_b},
\quad
\frac{D_{26}'}{\rho_a} = \frac{\lambda_a}{\lambda_a-\lambda_b} -2. 
\eeq
These identities may be obtained using partial fraction identities similar to those in eqs. \rf{3c1} and \rf{3c2}. 
Equations $\rf{cond}_2$ and  $\rf{cond}_3$ can  be rewritten 
\beqa{491}
\left[
\begin{array}{cc} 
D_{15}' & D_{25}'
\\
D_{16}' & D_{26}'-D_{16}'
\end{array}\right]
\left(
\begin{array}{c} 
2\nu \\ 1 \end{array}\right) = 
\left(
\begin{array}{c} 
0 \\ 0 \end{array}\right).
\eeqa
However, using \rf{490}, the determinant of the matrix is   
\beq{492}
D_{15}'D_{26}' - (D_{15}' +D_{25}')D_{16}'
= - 3 \rho_+\rho_-, 
\eeq
which is non-zero inside the triangle of figure \ref{fcube}. 
This gives us the important result:  there are no stationary values of $\nu$ inside the triangle of figure \ref{fcube}.  Hence, the only possible stationary values are on the edges.  
 
 \subsection{Stationary conditions on the triangle edges}
 
The analysis above for the three conditions \rf{cond} is not valid on the triangle edges in figure \ref{fcube}  because the quantities $\rho_\pm$ become zero and careful limits must be taken.  We avoid this route by considering the conditions \rf{cond} afresh for $\bf n$ directed along the three edges.   We find, as before, that $D_{14}' =0$ on the three edges, so that $\rf{cond}_1$  always holds.  Of the remaining two conditions, one is always satisfied, and imposing the other condition gives the answer sought.   

The direction $\bf n$ can be parameterized along each edge with a single variable.  Thus, ${\bf n} = 1p0$, $0\le p \le 1$, on edge 1.  Similarly, edges 2 and 3 are together covered by ${\bf n} = 11p$, with $0\le p < \infty$.  In  each case we also need to consider the two possible values of $\bf m$, which we proceed to do, focusing on the conditions $\rf{cond}_2$ and $\rf{cond}_3$.

\subsubsection{Edge 1:  ${\bf n} = 1p0$, $0\le p \le 1$, and ${\bf m} = p\bar{1}0$ or $001$}
 
For ${\bf m} = p\bar{1}0$ we find that $D_{15}'= D_{25}' = 0$, $D_{16}' = - D_{26}' = p-p^3$. 
Hence equation $\rf{cond}_2$  is automatically satisfied, while  equation $\rf{cond}_3$ becomes
\beq{311}
(\nu - 1) \, (p - p^3) = 0\, . 
\eeq
Conversely, for ${\bf m} = 001$ it turns out that $D_{16}'= D_{26}' = 0$, and $D_{15}' = - D_{25}' = p-p^3$.  In this case the only non-trivial equation from  equations \rf{cond} is the second one, 
\beq{312}
\nu \, (p - p^3) = 0\, . 
\eeq
Apart from the specific cases $\nu = 0$ or $\nu = 1$, 
equations  \rf{311} and \rf{312} imply that stationary values of $\nu$   occur only at the end points $p=0$ and $p=1$.  Thus, $\nu (001)$, $\nu (110, 1\bar{1}0)$, and $\nu (110, 001)$ are 
potential candidates for global extrema of $\nu$. 
 
\subsubsection{Edges 2 and 3:  ${\bf n} = 11p$, $0\le p < \infty $ and ${\bf m} = 1\bar{1}0$}

Proceeding as before  we find that $D_{16}'= D_{26}' = 0$,  $D_{15}' = \sqrt{2} p(1-p^2)/(2+p^2)^2$,   and $D_{25}' = p/[\sqrt{2}(2+p^2)] $.  Hence equation  $\rf{cond}_3$ is  automatically satisfied, while  equation $\rf{cond}_2$ becomes
\beq{321}
p\big[ (1-4\nu) p^2 + 2 + 4\nu\big] = 0\, . 
\eeq
The zero $p=0$ corresponds to ${\bf n} = 110$ which was considered above.  Thus, all three conditions \rf{cond} are met if $p$ is such that
 \beq{322}
p^2 = \big(\nu +1/2\big)/\big(\nu -1/4\big)\, . 
\eeq

Further progress is made using 
 the representation of equation \rf{sug} combined with the limiting values of $D$ which can be easily evaluated. 
 We find
\begin{subequations}\label{314}
\begin{align}\label{314a}
\frac{E_{111}}{E_{11p}} &= 1 +   \frac13 \big( \frac{1-p^2}{2+p^2} \big)^2\, E_{111} \chi \, , 
\\
\nu(11p\, , 1\bar{1}0)  \, \frac{E_{111}}{E_{11p}} &=  \nu_{111}-  \frac16 \big( \frac{1-p^2}{2+p^2} \big)\, E_{111} \chi \,  .
\label{314b}
\end{align}
\end{subequations}
Substituting for $p^2$ from equation \rf{322} into \rf{314} gives two coupled equations for $E_{11p}$ and $\nu(11p\, , 1\bar{1}0)$: 
\beq{324}
\frac{1}{E_{11p}} = \frac{1}{E_{111}} +   \frac{\chi}{48\nu^2} \, , 
\qquad
\frac{\nu }{E_{11p}} =  \frac{\nu_{111}}{E_{111}} +   \frac{\chi}{24\nu}  \,  .
\eeq
Eliminating $E_{11p}$ yields a single equation for possible stationary values of 
$\nu(11p\, , 1\bar{1}0)$: 
\beq{325}
\nu^2- \nu\, \nu_{111} - \frac{1}{48}E_{111} \chi = 0  \,  .
\eeq
We will return to this after considering the other possible $\bf m$ vector. 

\subsubsection{Edges 2 and 3:  ${\bf m}_- = pp\bar{2}$}

In this case $D_{15}'= D_{25}' = 0$,  $D_{16}' = \sqrt{2} p(1-p^2)/(2+p^2)^2$,   and $D_{26}' = p(p^2-4)/[\sqrt{2}(2+p^2)^2] $.  Equation $\rf{cond}_2$ holds, while  equation $\rf{cond}_3$ is zero if $p=0$, which is disregarded,  or  if $p$ is such that
 \beq{342}
p^2 = \big(\nu -3/2\big)/\big(\nu -3/4\big)\, . 
\eeq
The Young's modulus is independent of $\bf m$ and given by \rf{314a}, while $\nu$ satisfies 
\beq{314c}
\nu(11p\, , pp\bar{2}) \, \frac{E_{111}}{E_{11p}} =  \nu_{111} +  \frac{(1-p^2)(4-p^2)}{6 (2+p^2)^2} \, E_{111} \chi \,  .
\eeq
Using the value of $p^2$ from \rf{342} in equations  \rf{314a} and \rf{314c}  yields another pair of   coupled equations, for $E_{11p}$ and $\nu(11p\, , 001)$: 
\beq{344}
\frac{1}{E_{11p}} = \frac{1}{E_{111}} +   \frac{\chi}{48(\nu -1)^2} \, , 
\qquad
\frac{\nu }{E_{11p}} =  \frac{\nu_{111}}{E_{111}} +   \frac{\chi (\nu -\frac12)}{24(\nu -1)^2}  \,  .
\eeq
These imply a single equation for possible stationary values of 
$\nu(11p\, , 001)$: 
\beq{345}
(\nu -1)^2- (\nu -1)(\nu_{111}-1) - \frac{1}{48}E_{111} \chi = 0  \,  .
\eeq

\subsection{Definition of  $\nu_1$ and $\nu_2$}

The analysis for the three edges gives   a total  of seven candidates for global extrema:  $\nu (001)$, $\nu (110, 1\bar{1}0)$, and $\nu (110, 001)$ from the endpoints of edge 1,  and the four roots of equations \rf{325} and \rf{345} along edges 2 and 3.  The latter are very interesting because they are the only instances of possible extreme values associated with directions other than the principal directions of the cube (axes, face diagonals).   Results below will show that 
five of the seven candidates are global extrema, depending on the material properties.  These are $\nu (001)$, $\nu (110, 1\bar{1}0)$,  $\nu (110, 001)$ and the following two distinct roots of 
equations \rf{325} and \rf{345}, respectively,  
\begin{subequations}\label{222}
\beqa{222a}
\nu_1 &\equiv & \frac12 \nu_{111} - \frac12 \sqrt{ \nu_{111}^2 +  \frac16\big(\nu_{111}+1\big)
\big(\frac{\mu_1}{\mu_2}-1\big)   }, 
\\
\nu_2 & \equiv & \frac12 \big(\nu_{111} +1\big) + \frac12\sqrt{ \big(\nu_{111}-1\big)^2 +\frac16\big(\nu_{111}+1\big) \big(\frac{\mu_1}{\mu_2}-1\big)  } 
.  \label{222b}
\eeqa
\end{subequations}
The quantity $E_{111} \chi$ has been replaced  to emphasize the dependence upon the two parameters $\nu_{111}$ and the anisotropy ratio ${\mu_1}/{\mu_2}$. 
The associated directions follow from equations \rf{322} and \rf{342},
\begin{subequations}\label{223}
\begin{align}\label{223a}
\nu_1 &= \nu(11p_1\, , 1\bar{1}0), & p_1 = 
 \bigg(\frac{\nu_1+1/2}{\nu_1-1/4}\bigg)^{1/2}\, , 
\\ 
\nu_2 &= \nu(11p_2\, , p_2p_2\bar{2}), &  p_2 =
\bigg(\frac{\nu_2-3/2}{\nu_2 -3/4}\bigg)^{1/2}
 \, .  \label{223b}
\end{align}
\end{subequations}
A complete analysis is provided in Appendix B.  At this stage we note that 
$\nu_1$ is identical to the minimum value of $\nu$ deduced by Ting \&  Chen (2005), i.e.  equations (4.13) and (4.15) of their paper, with the minus sign taken in equation (4.13).  

\section{Material properties in terms of Poisson's ratios}  \label{sec5}

Results for the global extrema are presented after we introduce several  quantities. 

\subsection{Nondimensional parameters}

  It  helps to characterize the  Poisson's ratio in terms of two nondimensional material parameters  
  which we select as  $\nu_0$ and  $\chi_0$, where
 \beqa{e11}
\nu_0 &=& - \, {s_{12}}/{s_{11}} = (3\kappa- 2\mu_2)/(6\kappa+2\mu_2), 
\\
 \chi_0 &=&  \big( 2s_{11} - 2s_{12} - s_{44}\big) / s_{11} 
 = \frac{\mu_2^{-1} - \mu_1^{-1} }{(9\kappa)^{-1} +(3\mu_2)^{-1}} \, .  \label{e11b}
\eeqa
That is, $\nu_0$ is the axial Poisson's ratio $\nu (001,\cdot)$, independent of the orthogonal direction, and $\chi_0 = \chi/{s_{11}}$ is the nondimensional analogue of $\chi$.  
 Thus, 
\beq{e13}
 \nu ( {\bf n}, {\bf m} ) 
=\frac{\nu_0 - \frac12 \chi_0 D({\bf n}, {\bf m}) }{ 1 - \chi_0 F({\bf n})}  \, , 
\eeq
a  form which  shows clearly that  $ \nu $ is negative (positive) for all directions if   $\nu_0 <0$ and $\chi_0 >0$  ($\nu_0 >0$ and $\chi_0 <0$).    These conditions for cubic materials to be  completely auxetic (non-auxetic) were previously derived by Ting \&  Barnett (2005).   
The extreme values of the Poisson's ratio for a given $\bf n$ are 
\beq{z2}
\nu_\pm ({\bf n})=\frac{\nu_0 - \frac12 \chi_0 (F\pm H) }{ 1 - \chi_0 F}\, , 
\eeq
where $F$ is defined in \rf{df} and $H$ in \rf{r2a}. 
Thus,  $\nu_+$ is the minimum  (maximum) and $\nu_-$ the maximum (minimum)  if $\chi_0 >0$
( $\chi_0 <0$), respectively. 

The Poisson's ratio is a function of the direction pair $( {\bf n}, {\bf m} )$ and the material parameter pair $( \nu_0, \chi_0 )$, i.e. $\nu = \nu  ( {\bf n}, {\bf m},  \nu_0, \chi_0 )$.  The dependence upon $\nu_0$ has an interesting property:   for any orthonormal triad, 
\beq{gen}
 \nu ( {\bf n}, {\bf m} ,\nu_0 , \chi_0) +\nu ( {\bf n}, {\bf t} , 1-\nu_0 , \chi_0) =  1\, . 
\eeq
This follows from \rf{e13} and the identities \rf{fol}. Result \rf{gen} will prove useful later.

Several particular values of Poisson's ratio have been introduced: $\nu_0= {\nu }(001 ,\, {\bf m} ) $, $\nu_{111}={\nu }( 111,\, {\bf m} ) $ associated with the two directions $001$ and $111$ for which $\nu$ is independent of $\bf m$. These are  two vertices of the triangle in figure \ref{fcube}.   At the third vertex  (${\bf n} = 110$  along the face diagonals)  we have   
$\nu(  110,\, {\bf m} ) =  m_3^2 \nu_{001} +  (1-m_3^2)\nu_{1\bar{1}0}$ where, in the notation of  (Milstein \& Huang 1979), 
$\nu_{001}\equiv {\nu }( 110,\, 001 )$  and $\nu_{1\bar{1}0}\equiv{\nu }( 110,\, 1\bar{1}0 )$.  Three of these four values of Poisson's ratio associated with principal directions can be global extrema, and the fourth, $\nu_{111}$ plays a central role in the definition of $\nu_1$ and $\nu_2$ of \rf{222}. 
We therefore consider them in terms of the nondimensional parameters $\nu_0$ and $\chi_0$: 
\beq{4val}
\nu_{111} =  \frac{ \nu_0 - \frac16  \chi_0  }{ 1 -\frac13  \chi_0 }\, ,
\qquad
\nu_{001}= \frac{ \nu_0  }{ 1 -\frac14  \chi_0 }\, ,
\qquad
\nu_{1\bar{1}0}  = \frac{ \nu_0 -\frac14  \chi_0  }{ 1 -\frac14  \chi_0 }\, .
\eeq
We return to $\nu_1$ and $\nu_2$ later. 

\begin{figure}[ht]
				\begin{center}	
				\includegraphics[width=4.0in , height=2.8in 					]{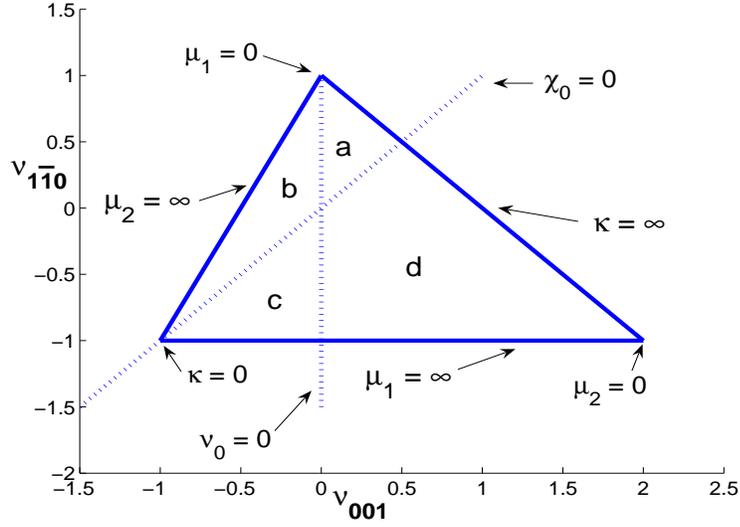} 
	
	\caption{The interior of the triangle in the $\nu_{001},\, \nu_{1\bar{1}0}$ plane represents the entirety of possible cubic materials with positive definite strain energy.  The vertices  correspond to $\kappa=0$,  $\mu_1=0$ and  $\mu_2=0$, as indicated.  The edges of the triangle opposite the  vertices  are the limiting cases in which $\kappa^{-1}$,  $\mu_1^{-1}$ and  $\mu_2^{-1}$ vanish, respectively.   The dashed curves correspond to $\nu_0=0$ (vertical) and $\chi_0=0$ (diagonal) and the regions $a$, $b$, $c$ and $d$ defined by these lines coincide with the four cases in equation \rf{abcd} respectively. }
		\label{ftriangle} \end{center}  
	\end{figure}

\subsection{Positive definiteness and Poisson's ratios}

In order to   summarize the global extrema on $\nu$ we first need to   consider the range of possible material parameters. 
It may be shown that  the requirements for the strain energy to be positive definite:  $\kappa >0$, $\mu_2 >0$ and $\mu_1 >0$, can be expressed in terms of  $\nu_0$ and $\chi_0$ as  
\beq{e12}
-1 < \nu_0 < 1/2, \qquad \chi_0 < 2(1+\nu_0)\, .   
\eeq
It will become evident that the global extrema for $\nu$ depend most simply on the two values for $\bf n$ along a face diagonal: $\nu_{001}$ and $\nu_{1\bar{1}0}$.   The constraints \rf{e12} become
\beq{wq3}
-1 < \nu_{1\bar{1}0} < 1 , \qquad -\frac12 (1-\nu_{1\bar{1}0} ) < \nu_{001} < 1-\nu_{1\bar{1}0}
\, ,   
\eeq
which  define  the interior of a triangle in the $\nu_{001},\, \nu_{1\bar{1}0}$ plane, see figure \ref{ftriangle}.  This figure also indicates the lines $\nu_0=0$ and $\chi_0=0$ (isotropy).  It may be checked that the 
 four quantities $\{\nu_0,\, \nu_{111},\, \nu_{001},\, \nu_{1\bar{1}0}\}$ are different  as long $\chi_0 \ne 0$, with the exception of 
$\nu_{001}$ and $\nu_0$ which are distinct  if  $\nu_0 \chi_0 \ne 0$.  
Consideration of  the four possibilities yields the ordering  
\begin{subequations} \label{abcd}
\beqa{abcda}
 & \ \  0 \, <\, \nu_{001} \, < \, \nu_0  \, < \, \nu_{111} \, < \, \nu_{1\bar{1}0} \, < \,1\, 
& \qquad \mbox{for  }  \nu_0 >0 , \,\chi_0 <0 ,\,
\\ 
& -1\,<\, \nu_0  \, < \, \nu_{001} \, < \, \nu_{111} \, < \, \nu_{1\bar{1}0}\, < \,0 \, 
& \qquad \mbox{for  }  \nu_0 <0 , \, \chi_0 <0 \, , \label{abcdb}
 \\ 
& -1\,<\, \nu_{1\bar{1}0} \, < \,  \nu_{111} \,  < \, \nu_{001} \, < \, \nu_0\, < \,0   \,
 &\qquad \mbox{for  }  \nu_0 <0 , \,\chi_0 >0 ,\, \label{abcdc}
\\ 
&\ \ 0\, <\, \nu_{1\bar{1}0} \, < \,  \nu_{111} \, < \,  \nu_0  \, < \, \nu_{001} \,< \,2 \, 
 & \qquad \mbox{for  }  \nu_0 >0 , \,\chi_0 >0 .\label{abcdd}
\eeqa
\end{subequations}
Note that $\nu_{111}$ is never a maximum or minimum. 
We will see below that  \rf{abcda} is the only  case for which  the extreme values coincide with the global extrema for $\nu$.  This is one of the reasons the classification of the extrema for $\nu$ is relatively  complicated, requiring that we identify several distinct values.  In particular, the global extrema depend upon more than $\sgn \nu_0$ and  $\sgn \chi_0$, but are best characterized by  the  two independent nondimensional parameters  $\nu_{001}$ and $\nu_{1\bar{1}0}$.

We are now ready to define the global extrema.

\section{Minimum and maximum Poisson's ratio}\label{sec6}

\begin{figure}[ht]
				\begin{center}	
				\includegraphics[width=4.5in , height=2.8in 					]{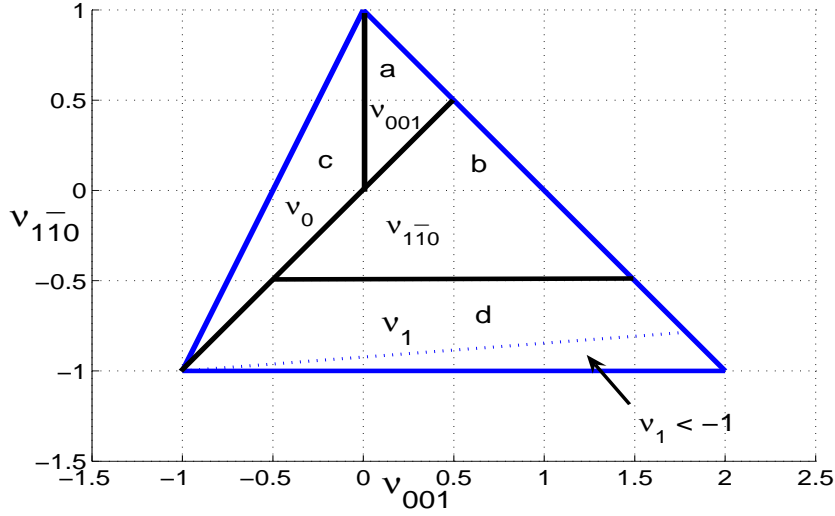} 
	
	\caption{The global minimum  of Poisson's ratio based on table 1.   The value of $\nu_{\rm min}$ depends upon the location of 	the cubic material parameters in the four distinct regions a, b, c and d, defined by the heavy lines inside the triangle of possible materials. The diagonal dashed line delineates the region in which $\nu_{\rm min}<-1$, from eq. \rf{st}.}
		\label{fmin} \end{center}  
	\end{figure}

Tables 1 and 2 list the values of the global minimum $\nu_{\rm min}$ and the global maximum $\nu_{\rm max}$, respectively, for  all  possible combinations of elastic parameters.  
For table 1, $\nu(001)$, $\nu_{001}$ and $ \nu_{1\bar{1}0}$ are defined in \rf{4val}, and  $\nu_1$ and $p_1$ are defined in \rf{222a} and \rf{223a}.  For table 2,   $\nu_2$ and $p_2$  are defined in \rf{222b} and \rf{223b}.  No second condition is necessary to define the region for case e, which is clear from  figure \ref{fmax}.
The data in tables 1 and 2 are illustrated in figures \ref{fmin} and \ref{fmax}, respectively, which  define the global extrema for every point in the interior of the  triangle defined by \rf{wq3}.   The details of the analysis and related numerical tests leading to these results are presented in Appendix B.


\begin{table}		
\begin{center}	
\caption{The global minimum of Poisson's ratio for cubic materials}  
\begin{tabular}{rccccc} \hline  
  $\,\,\nu_{\rm min}$ \,\,   & \,\,$\bf n$\,\, & \,\, $\bf m$  \,\, &
  \,  condition 1 \,&  \, condition 2  \, & Fig. \ref{fmin}
  \\ \hline 
 $0< \nu_{001}$ & $110$ & $001$ & $\nu_{001}>0$  & $\nu_{1\bar{1}0} > \nu_{001}$ & a \\
 $-\frac12 < \nu_{1\bar{1}0}$ & $110$ & $1\bar{1}0$ &  $\nu_{1\bar{1}0} > -\frac12$  & $\nu_{1\bar{1}0} < \nu_{001}$ & b \\
 $-1< \nu_0$~~ & $001$ & arbitrary &$\nu_{001}<0$  & $\nu_{1\bar{1}0} > \nu_{001}$ & c  \\
  $-\infty < \nu_1$~~ & \, $11p_{_1}$ & $1\bar{1}0$ & $\nu_{1\bar{1}0} < -\frac12$  & $\nu_{1\bar{1}0} < \nu_{001}$ &  d\\
\hline 		 
\end{tabular}
\end{center}
\end{table}

\begin{table}	
\begin{center}			
\caption{The global maximum of Poisson's ratio}
\begin{tabular}{lccccc} \hline  
  $\,\,\nu_{\rm max}$ \,\,   & \,\,$\bf n$\,\, & \,\, $\bf m$ & \,  condition 1 \,&  \, condition 2  \, & Fig. \ref{fmax} \,\,
  \\ \hline 
~~~$\nu_1 < -\frac12$ & $11p_{_1}$ & $1\bar{1}0$ &  $\nu_{1\bar{1}0} < -\frac12$ & $\nu_{1\bar{1}0} > \nu_{001}$ &  a\\
~~ $\nu_0 <0 $ & $001$ & arbitrary & $\nu_{001}<0$  & $\nu_{1\bar{1}0} < \nu_{001}$ &  b \\
 $\nu_{1\bar{1}0} < 1$ & $110$ & $1\bar{1}0$ & $\nu_{1\bar{1}0} > -\frac12$ & $\nu_{1\bar{1}0} > \nu_{001}$ & c \\
 $\nu_{001} < \frac32 $ & $110$ & $001$ & $0 <\nu_{001}< \frac32$  & $\nu_{1\bar{1}0} < \nu_{001}$ &  d \\
~~$\nu_2 <\infty$ &  \, $11p_{_2}$ & $p_{_2}p_{_2}\bar{2}$ & $\nu_{001} > \frac32$  &     &   e\\
\hline
\end{tabular}
\end{center}
\end{table}	

\subsection{Discussion}\label{Discussion}

Conventional wisdom prior to Ting \&  Chen (2005) was that the extreme values were characterized by the face diagonal values $\nu_{001}$ and $\nu_{1\bar{1}0}$.  But as equation \rf{abcd} indicates, even these are not always  extrema, since $\nu_0 = \nu(001,{\bf m}) $ can be maximum or minimum under appropriate circumstances (equations \rf{abcdc} and \rf{abcdd}, respectively).  The extreme values in equation \rf{abcd} are all bounded by the limits of the triangle in figure \ref{ftriangle}.  Specifically, they limit the Poisson's ratio to lie between $-1$ and $2$.  Ting \&  Chen (2005) showed by explicit demonstration that this is not the case, and that values less than $-1$ and larger than $2$ are feasible, and remarkably,  no lower or upper limits  exist for $\nu$.    

\begin{figure}[ht]
				\begin{center}	
				\includegraphics[width=4.5in , height=2.8in 					]{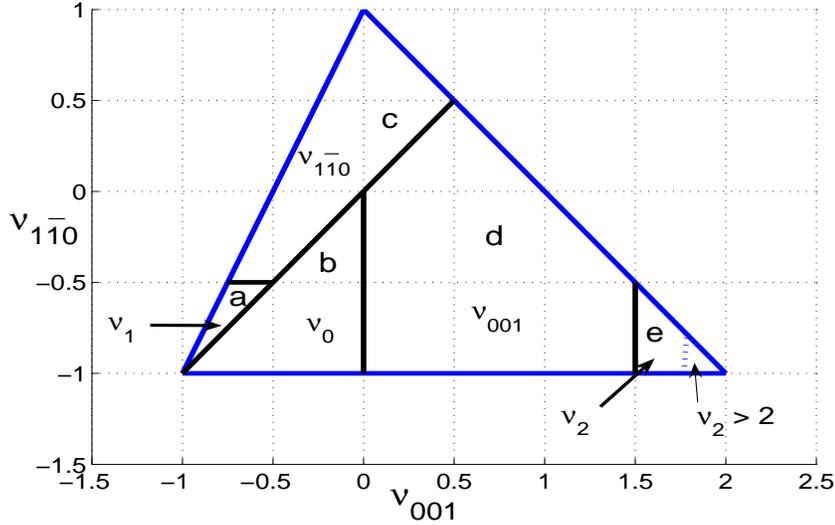} 
	
	\caption{The global maximum  of Poisson's ratio based on table 2.   The value of $\nu_{\rm max}$ depends upon the location of $(\nu_{001},\, \nu_{1\bar{1}0})$ in  five distinct regions defined by the heavy lines. The  dashed line delineates the (small) region in which $\nu_{\rm max}>2$, from eq. \rf{st}.}
		\label{fmax} \end{center}  
	\end{figure}

The Ting \&  Chen ``effect" occurs in figure \ref{fmin} in the region where 
$\nu_{\rm min}=\nu_1$ and in figure  \ref{fmax} in the region  $\nu_{\rm max}=\nu_2$.  	
Using equation \rf{222a} we can determine that  $\nu_{\rm min}$ is strictly less than $-1$ if 
$(\mu_1/\mu_2-1) > 24$. 
Similarly, equation  \rf{222b} implies that $\nu_{\rm max}$ is strictly greater than $2$ 
if $(\mu_1/\mu_2-1) (\nu_{111} +1 ) >24 (2-\nu_{111})$.
By converting these inequalities we deduce 
\begin{subequations}\label{st}
\begin{align}\label{sta}
&\nu_{\rm min} < -1 
\  &\Leftrightarrow &\qquad  \mu_2 < \frac{\mu_1}{25} 
\  &\Leftrightarrow &\qquad   \nu_{001} -13 \nu_{1\bar{1}0} >12, &  
 \\
&\nu_{\rm max} > 2 
\  & \Leftrightarrow &\qquad
\mu_2 < \bigr( \frac{25}{\mu_1} + \frac{16}{\kappa} \bigr)^{-1} 
\ & \Leftrightarrow &\qquad  13\nu_{001} -  \nu_{1\bar{1}0} > 24 .&
\end{align}
\end{subequations}
The two subregions defined by the $\nu_{001}$, $\nu_{1\bar{1}0}$ inequalities are depicted in figures \ref{fmin} and \ref{fmax}.  They define  neighbourhoods of the $\mu_2 = 0$  vertex, i.e.   $(\nu_{001}, \nu_{1\bar{1}0}) = (2,-1)$, where the extreme values of $\nu$ can achieve arbitrarily large positive and negative values.  
The condition for $\nu_{\rm min} < -1$   is independent of the bulk modulus $\kappa$.  Thus, the occurrence of  negative values of $\nu$  less than $-1$ does not necessarily imply that relatively large positive values (greater than $2$) also occur, but the converse is true.  This is simply a consequence of the fact that the dashed region  near the tip $\mu_2=0$ in figure \ref{fmax} is contained  entirely within the dashed region of figure \ref{fmin}. 

These results indicate that the necessary and sufficient condition for the occurrence of large extrema for $\nu$ is that $\mu_2$ is much less than either $\mu_1$ or $\kappa$.   $\mu_2$ is either the maximum or minimum of $G$, and it is associated with directions pairs along orthogonal face diagonals, $\mu_2 = G(110,\, 1\bar{1}0)$. 
Hence, the Ting \&  Chen effect requires that this shear modulus is much less than $\mu_1 = G(001,\, {\bf m})$, and much less than the bulk modulus $\kappa$. 
In the limit of very small $\mu_2$   equations \rf{222} give 
$\nu_{1, \, 2} \approx   \mp 
\sqrt{ (\nu_{111} +1) \mu_1/(24 \mu_2)}$. 
Ting (2004) found that the extreme values are $\nu \approx\pm \sqrt{3/(16\delta)}+$O$(1)$  for small values of their parameter $\delta$.  In current notation this is 
$\delta = 9/[1 + E_{111}\, \chi ]$, and replacing $E_{111}\, \chi $ the two theories are seen to  agree.

The implications of small $\mu_2$ for Young's modulus are apparent.  Thus, $E_{\rm min}/E_{\rm max}$ $=$O$(\mu_2/\mu_1)$, and equation $\rf{sug}_1$ indicates that $E({\bf n})$ is small everywhere except near the $111$ direction, at which it reaches a sharply peaked maximum.  
Cazzani \&  Rovati (2003) provide numerical examples illustrating the directional variation of $E$ for a  range of auxetic and non-auxetic cubic materials, some of which are considered below.  Their  3-dimensional plots of $E({\bf n})$ for materials with very large values of $\mu_1/\mu_2$ (see Table 3 below) look like very sharp starfish.   
Although the directions at which $\nu_1$ and $\nu_2$ are large in magnitude are close to the $111$ direction,  the value of $E$ in the stationary directions can be quite different  from $E_{111}$.  The precise values  
of the Young's modulus, $E_{11p_{_1}}$ and $E_{11p_{_2}}$,  at the associated stretch directions are  given by the  identities:
\beq{411}
 \frac{E_{111} }{E_{11p_{_1}}} + \frac{\nu_{111} }{\nu_1} = 2, 
\qquad 
\frac{E_{111} }{E_{11p_{_2}} } + \frac{\nu_{111} -1}{\nu_2 - 1}  = 2\, . 
\eeq
The first identity follows from the pair of equations \rf{324} and the second from \rf{344}.  Equations \rf{411} indicate that if $\nu_1$  or $\nu_2$ become large in magnitude then the second  term in the left member is negligible, and  the associated value of the Young's modulus is approximately one half of the value in the $111$ direction. Thus, large values of $|\nu|$ occur in directions at which $E \approx \frac12 E_{111}$.  Such directions, by their nature, are close to $111$.

We note  that $\nu_1$ appears in both figures \ref{fmin} and \ref{fmax}.  The double occurrence is not  surprising if one considers that $\nu_{001}$,  $\nu_{1\bar{1}0}$ and $\nu_0 $ also occur in both the minimum and maximum.  It can be checked that in the region where  $\nu_1$ is the maximum value in  figure  \ref{fmax}  it satisfies 
$-1 < \nu_1 < -1/2$. In fact it is very close to but not equal to $\nu_{1\bar{1}0}$ in this region, and numerical results indicate that $|\nu_1 - \nu_{1\bar{1}0}| < 4\,\times 10^{-4}$ in this small sector.  

What is special about the transition values in figures \ref{fmin} and \ref{fmax}: 
$\nu_{001} =3/2$ and $\nu_{1\bar{1}0}= -1/2$? Quite simply, they are the values of $\nu_1$ and $\nu_2$ as the stationary directions $\bf n$ approach the face diagonal direction $110$. Thus, $\nu_1$ and $\nu_2$ are both  the continuation of the face diagonal value 
$\nu_{1\bar{1}0}$, but on two different branches. See Appendix B for further discussion.

\begin{figure}[ht]
				\begin{center}	
				\includegraphics[width=4.5in , height=3.0in 					]{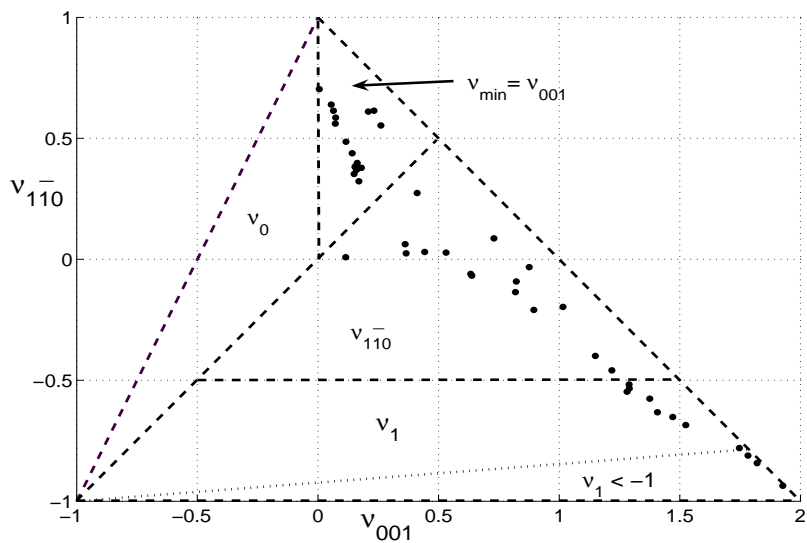} 
	\caption{The 44 materials considered are indicated by dots on the chart showing the  $\nu_{\rm min}$ regions, see figure \ref{fmin}.  }
		\label{fig6} \end{center}  
	\end{figure}

\subsection{Application to cubic materials}\label{numbers}

We conclude by considering elasticity data for 44 materials with cubic symmetry, figure \ref{fig6}.  
The data are from Musgrave (2003) unless  otherwise noted. 
The cubic materials in the region where $\nu_{\rm min} = \nu_{001}$ are as follows, with the coordinates $(\nu_{001},\nu_{1\bar{1}0})$ for each:
GeTe–SnTe\footnote{Data from Landolt and Bornstein (1992), see also Cazzani and Rovati (2003).} 
(mol\% GeTe=0)	(0.01,	0.70), 
RbBr\footnotemark[1]	(0.06,	0.64), 
KI	(0.06,	0.61), 
KBr	(0.07,	0.59), 
KCl	(0.07,	0.56), 
Nb\footnotemark[1]	(0.21,	0.61), 
AgCl	(0.23,	0.61), 
KFl	(0.12,	0.49), 
CsCl	(0.14,	0.44), 
AgBr	(0.26,	0.55), 
CsBr	(0.16,	0.40), 
NaBr	(0.15,	0.38), 
NaI	(0.15,	0.38), 
NaCl	(0.16,	0.37), 
Cr–V\footnotemark[1] (Cr–0.67 at.\% V)	(0.15,	0.35),
CsI	(0.18,	0.38),
NaFl	(0.17,	0.32).  
This lists them roughly in the order from top left to lower right.  Note that  all the materials considered  have positive $\nu_{001}$.  The materials with    
 $\nu_{\rm min} = \nu_{001}$ also have  $\nu_{\rm max} = \nu_{1\bar{1}0}$, so the coordinates of the above materials  correspond to their extreme values of $\nu$.  The extreme values are also given by the coordinates  in the region with 
$\nu_{\rm min} = \nu_{1\bar{1}0}$,  $\nu_{\rm max} = \nu_{001}$.  The materials there are: 
Al	(0.41,	0.27),
diamond	(0.12,	0.01),
Si	(0.36,	0.06),
Ge	(0.37,	0.02),
GaSb	(0.44,	0.03),
InSb	(0.53,	0.03),
 Cu–Au\footnotemark[1]	(0.73,	0.09),
Fe	(0.63,	-0.06),
Ni	(0.64,	-0.07),
Au	(0.88,	-0.03),
Ag	(0.82,	-0.09),
Cu	(0.82,	-0.14),
$\alpha$-brass	(0.90,	-0.21),
 Pb\footnotemark[1]	(1.02,	-0.20),
 Rb\footnotemark[1]	(1.15,	-0.40),
 Cs\footnotemark[1]	(1.22,	-0.46).

\begin{table}
\begin{center}
 \caption{Properties of the 11 materials of cubic symmetry in figure \ref{fig6} with 
 $\nu_{1\bar{1}0}< -1/2$.  The boldfaced numbers indicate $\nu_{\rm min}$ and $\nu_{\rm max}$.  
 Unless otherwise noted the data are from Landolt and Bornstein (1992).  G\&S indicates Gunton and Saunders (1975).} 
\begin{tabular}{lrrcrrcc} \hline  
  Material   &      $\nu_{001}$ & $\nu_{1\bar{1}0}$ &   $\nu_1$     &  $p_1$ &  $\nu_2$~     &  $p_2$ & ${\mu_1}/{\mu_2}$ 
  \\ 
  \hline 
 $\beta$-brass (Musgrave 2003) & {\bf 1.29}	&	 -0.52	&	{\bf -0.52}	&	0.15	&		&	 	&	8.5 \\
 Li	 & {\bf 1.29}	&	-0.53	&	{\bf -0.54}	&	0.21	&	 	&	 	&	8.8 \\
 Al–Ni (at 63.2\% Ni and at 273 K) & {\bf 1.28}	&	-0.55	&	{\bf -0.55}	&	0.25	&		&	 	&	9.1 \\
 Cu–Al–Ni (Cu–14\% Al–4.1 \% Ni) & {\bf 1.37}	&	 -0.58	&	{\bf -0.59}	&	0.32	&		&	 	&	10.2 \\
 Cu–Al–Ni (Cu–14.5\% Al–3.15\% Ni) & {\bf 1.41}	&	 -0.63	&	{\bf -0.66}	&	0.42	&		&	 	&	12.1 \\
 Cu–Al–Ni	 & {\bf 1.47}	&	-0.65	&	{\bf -0.69}	&	0.45	&		&	 	&	13.1 \\
 Al–Ni (at 60\% Ni and at 273 K) & 1.53	&	-0.68	&	{\bf -0.74}	&	0.50	&	{\bf 1.53}	&	0.18	&	15.0 \\
 In–Tl (at 27\% Tl, 290K) (G\&S)	 & 1.75	&	-0.78	&	{\bf -0.98}	&	0.62	&	{\bf 1.89}	&	0.59	&	24.0 \\
 In–Tl (at 28.13\% Tl)	 & 1.78	&	-0.81	&	{\bf -1.08}	&	0.66	&	{\bf 2.00}	&	0.63	&	28.6 \\
 In–Tl (at 25\% Tl)	 &1.82	&	-0.84	&	{\bf -1.21}	&	0.70	&	{\bf 2.14}	&	0.68	&	34.5 \\
 In–Tl (at 27\% Tl, 200K) (G\&S)	 & 1.93	&	-0.94	&	{\bf -2.10}	&	0.83	&	{\bf 3.01}	&	0.82	&	90.9 \\
\hline
\end{tabular}
\end{center}
\end{table}


Materials with $ \nu_{1\bar{1}0} < -1/2$ are listed in table 3.   
These  all lie within the region where the minimum is $\nu_1$, and  of these,  five materials are in the sub-region where the maximum is $\nu_2$.  Three materials
 are in the sub-regions with $\nu_1< -1$  and $\nu_2>2$. 
These Indium Thallium alloys  of different composition and at different temperatures  are close to the stability limit where they undergo a martensitic phase transition from face-centered cubic form to  face-centered tetragonal.  The transition is discussed by, for instance, Gunton and Saunders (1975), who also provide data on another even more auxetic sample: In–Tl (at 27\% Tl, 125K). This material 
is so close to the $\mu_2=0$ vertex,  with  $\nu_{001}= 1.991$,  $\nu_{1\bar{1}0}= -0.997$ and 
$\mu_1/\mu_2 = 1905$ (!) that we do not include it in the table or the figure for being too close to the phase transition, or equivalently, too unstable (it has $\nu_1 = -7.92$ and   $\nu_2=   8.21$). 

We note that the stretch directions for the extremal values of $\nu$, defined by ${\bf n} = 11p_1$ and ${\bf n} = 11p_2$, are distinct.  As the materials approach the $\mu_2=0$ vertex the directions coalesce as they tend towards the cube diagonal $111$.  
The three materials in table 3 with $\nu_{\rm min}<-1$ and $\nu_{\rm max}> 2$ are close to the incompressibility limit, the line $\kappa = \infty$ in figure \ref{ftriangle}.  In this limit  both the  cube diagonal and axial Poisson's ratios tend to $1/2$, i.e. 
$\nu_{111}= \nu_0 = 1/2$, and  
\beq{068}
\nu_1 = \frac14 - \frac14 \sqrt{\frac{\mu_1}{\mu_2}},
\quad 
\nu_2= \frac34 + \frac14 \sqrt{\frac{\mu_1}{\mu_2}},
\quad 
p_1=p_2 = \sqrt{1 - 3  \sqrt{\frac{\mu_2}{\mu_1}}}, 
\qquad  
 \kappa \rightarrow \infty .  
\eeq
These  are reasonable approximations for the last three materials in table 3, which clearly satisfy $\nu_1 + \nu_2 \approx 1$, and $p_1 \approx p_2$.

\section{Summary}\label{Summary}

Figures \ref{fmin} and \ref{fmax} along with tables 1 and 2 are the central  results which summarize the extreme values of Poisson's ratio for all possible values of the elastic parameters for solids with positive strain energy and cubic material symmetry.  The application of the related formulas to the materials in figure \ref{fig6} shows that values less than $-1$ and greater than $+2$ are associated with certain stretch directions in some Indium Thallium alloys. 

 
\begin{acknowledgements}  Discussions with Prof. T. C. T. Ting are  appreciated.  \end{acknowledgements}

\appendix{Extreme values of $D({\bf n},{\bf m})$ for a given ${\bf n}$}

The extreme values of $D( {\bf n}, {\bf m})$ as a function of ${\bf m}$ for a given direction ${\bf n}$ can be determined using Lagrange multipliers $\lambda ,\, \rho$, and the generalized function 
\beq{dd}
f({\bf m} ) = D({\bf n}, {\bf m} )- \lambda |{\bf m}|^2 - 2\rho{\bf n}\cdot{\bf m} \, .
\eeq
Setting to zero the partial derivatives of $f$ with respect to $m_1$, $m_2$, $m_3$, implies
three equations, which may be solved to give 
\beq{im2}
{\bf m}=   \big(
\frac{\rho n_1}{n_1^2 - \lambda},\, \frac{\rho n_2}{n_2^2 - \lambda},\,\frac{\rho n_3}{n_3^2 - \lambda} \big),
\eeq
where  $\lambda ,\, \rho$ follow from the constraints ${\bf n}\cdot{\bf m}=0$ and $|{\bf m}|^2 =1$.  These are, respectively, \rf{101} and 
\beq{im3b}
\bigg[
\frac{n_1^2}{(n_1^2 - \lambda)^2} +\frac{n_2^2}{(n_2^2 - \lambda)^2}+\frac{n_3^2}{(n_3^2 - \lambda)^2}
 \bigg]\, \rho^2 = 1\, .  
\eeq
Equation \rf{101} implies that $\lambda$ is a root of the quadratic equation \rf{mm3}  
and \rf{im3b} yields the normalization factor $\rho$.  These results are summarized in equations \rf{sum} and \rf{dmm}. 

It may be easily checked that the generalized function $f$ is zero at the extremal values of $D$.  But $f = D-\lambda$, and hence the extreme values of $ D({\bf n}, {\bf m} )$ are simply the two roots of the quadratic \rf{mm3}, $0 \le \lambda_- \le \lambda_+ \le 1/2$.  Note that the extreme values depend only upon the invariants of the tensor ${\bf M}$ with components $M_{ij} = D_{ijkl}n_kn_l$.  Although this is a second order tensor and normally possesses three independent invariants, one is trivially a constant: tr${\bf M}=1$. The others are, e.g. tr${\bf M}^2=n_1^4+n_2^4+n_3^4=1- 2F({\bf n})$ (see equation \rf{fol}) and  det${\bf M}=n_1^2n_2^2n_3^2$. 

The above formulation is valid as long as 
$(n_1^2 - n_2^2)(n_2^2 - n_3^2)(n_3^2 - n_1^2) \ne 0$. 
For instance, if $n_2^2 = n_1^2$, then $\lambda_-$, $\lambda_+=$ min, max $(n_1^2, 3n_1^2n_3^2)$.  The ${\bf m}$ vector associated with $\lambda= n_1^2$ is undefined, according to \rf{im2}.  However, by taking the limit  $n_2^2 \rightarrow n_1^2$ it can be shown that ${\bf m} \rightarrow \pm  (1,-1,0)/\sqrt{2}$.  The other vector corresponding to  $\lambda= 3n_1^2n_3^2$ has no such singularity, and is ${\bf m} = \pm (n_3, n_3, -2n_1)/\sqrt{2}$.

The identity \rf{3c2} may be obtained by noting that each term can be split, e.g. 
$n_1^4/(n_1^2 - \lambda ) = n_1^2 + \lambda/(n_1^2 - \lambda )$,  then using the fundamental 
relation  \rf{101} with $n_1^2 +n_2^2+n_3^2=1$.  Various other identities can be found, e.g. 
\beq{other}
\frac{n_1^6}{(n_1^2 - \lambda_+)(n_1^2 - \lambda_-)} +
\frac{n_2^6}{(n_2^2 - \lambda_+)(n_1^2 - \lambda_-)} +
\frac{n_3^6}{(n_3^2 - \lambda_+)(n_1^2 - \lambda_-)} 
 = 1\, .
 \eeq

\appendix{Analysis}


Here we derive stationary conditions for directions $\bf n$ along the edges of the triangle in figure \ref{fcube} by direct analysis.  Numerical tests are performed for the entire range of material parameters.  The results  are consistent with and reinforce those of \S\ref{sec4}. 

The limiting Poisson's ratios of \rf{z2} are expressed $\nu_\pm ({\bf n}) \equiv \nu_\pm (\alpha ,\beta)$ in terms of two numbers, 
 where 
\beq{z6}
n_1^2 = (1+\alpha) \big(\frac13 - \beta\big) , \qquad
n_2^2 = (1-\alpha) \big(\frac13 - \beta\big),\qquad n_3^2 = \frac13 +2 \beta . 
\eeq
The range of $(\alpha, \,  \beta )$ which  needs to be considered  is $
0\le \alpha \le 1$, $\frac{\alpha}{3(3+\alpha)} \le \beta \le 1/3$, corresponding to the triangle in figure \ref{fcube}. 
This parameterization allows quick numerical searching for global extreme values of $\nu$ for a given cubic material.  

We first consider  the three edges  as shown in figure \ref{fcube} in turn. 
Edge 1 is defined by $\alpha = 1$, $1/12 \le \beta \le 1/3$. The limiting values are 
$\nu_- (1,\beta)={\nu_0  }/\big[1 - $$\big(\frac13 -  \beta  \big)\big(\frac13 +2\beta \big)2\chi_0 \big]$ and 
$\nu_+  (1,\beta) =1 - \nu_- (1,\beta) (1-\nu_0 )/\nu_0$.     
The extreme values are obtained at the ends: 
$\nu_- (1,1/12) = \nu_{001}$, 
$\nu_+ (1,1/12) = \nu_{1\bar{1}0}$, 
$\nu_- (1,1/3) =\nu_+ (1,1/3) = \nu_0$. 
These possible global extreme values agree with those of \S\ref{sec4}. 

Inspection of  figure \ref{fcube} shows that edges 2 and 3 can be considered by looking at $\nu_\pm(0,\, \beta)$ for $-1/6\le \beta \le 1/3$.
Straightforward calculation gives 
\beq{t2}
\nu_-(0,\beta)  = \frac{\nu_0 - \frac12 (\frac13 - \beta) \chi_0}{1 - (\frac13 - 3\beta^2) \chi_0},
\qquad
\nu_+(0,\beta)  = \frac{\nu_0 - \frac12 (\frac13 - \beta) (1+ 6\beta)\chi_0}{1 - (\frac13 - 3\beta^2) \chi_0}. 
\eeq
A function of the form $f/g$ is stationary at  $f/g = f'/g'$.  Applying this to the expressions in \rf{t2} implies that the extreme values of $\nu_-$ and $\nu_+$ satisfy, respectively,  
\beq{t6}
\nu_-(0, \beta )  = \frac{1}{12 \beta }\, ,  
\qquad 
\nu_+(0, \beta ) = 1 - \frac{1}{12 \beta }\, . 
\eeq
Combining equations \rf{t2} and \rf{t6} gives in each case a quadratic equation in $\beta$.  Thus, the extreme values of $\nu_- $ and $\nu_+ $ are at  $\beta = \beta_{-\pm}$ and $\beta = \beta_{+\pm}$,  the roots of the quadratic equations.  
The first identity, $\rf{t6}_1$ was found by Ting \&  Chen (2005), their equation (4.15). 

To summarize the analysis for the  three edges: 
Extreme values of Poisson's ratio on the 3 edges are at the ends of edge 1, and on edges 2 and 3  given by $\nu_\pm$ of equations \rf{t2}-\rf{t6}.

\subsection{Numerical proof of tables 1 and 2}

A numerical test was performed over the range of possible 
 materials.  This required searching the entire two-dimensional range for $\alpha,\, \beta$.   Consideration of all possible materials then follows by allowing the material point to range throughout the triangle of figure \ref{ftriangle}.  
In every case it is  found  that the extreme values of $\nu$ occur on the edge of the irreducible $1/48$th element of the cube  surface.   Furthermore, the extreme values are never found to occur along  edge 2.  
Extreme values on edge 3 in figure \ref{fcube} can be found by considering edge 3' instead, i.e. $\alpha = 0$, $-1/6 < \beta < 0$.  This implies as possible extrema one of $\nu_- (0, \beta_{-\pm})$ and one of $\nu_+ (0, \beta_{+\pm})$. 
 We define these as 
$\nu_1'= \nu_- (0, \beta_{--})$, 
and $\nu_2' = \nu_+ (0, \beta_{+-})$, where the signs correspond to the sign of the discriminant in the roots, then   they  are given explicitly  as
\begin{align}\label{gmin2}
\nu_1' &= \frac{-1}{2\big( 1 - \frac{\chi_0}{3}\big) }
\bigg\{
\big(  \frac{\chi_0}{6} - \nu_0\big) + \bigg[ \big(  \frac{\chi_0}{6} - \nu_0\big)^2
+ \frac{\chi_0}{12}\big( 1 - \frac{\chi_0}{3}\big) \bigg]^{1/2}\bigg\},
\\
\nu_2' &= \frac{1}{2\big( 1 - \frac{\chi_0}{3}\big) }
\bigg\{
\big(  1+ \nu_0 - \frac{\chi_0}{2} \big) + \bigg[ \big( 1-\nu_0 -\frac{\chi_0}{6} \big)^2
+ \frac{\chi_0}{12}\big( 1 - \frac{\chi_0}{3}\big) \bigg]^{1/2}\bigg\}\, . 
\label{gmax}
\end{align}
It may be checked that $\nu_1'=\nu_1$ and $\nu_2'=\nu_2$, in agreement with equation \rf{222}. 

The numerical results indicate the potential extrema come from the five values: 
$\nu_0$, 
$\nu_{001}$, 
$\nu_{1\bar{1}0}$,  
$\nu_1$ and 
$\nu_2$. 
It turns out that each is an extreme  for some range of material properties. Thus, the first four are necessary to define the global minimum, see table 1 and figure \ref{fmin}, while all five occur in the description of the global maximum,  in table 2 and figure \ref{fmax}.

Although a mathematical proof has not been provided for the veracity of tables 1 and 2, and figures  \ref{fmin} and \ref{fmax}, it is relatively simple to do a numerical test, {\it a posteriori}. By performing the numerical search as described above, and subtracting the extreme values of tables 1 and 2, one finds zero, or its numerical approximant for all points in the interior of the triangle of possible materials, figure \ref{fcube}.  

In order to further justify the results as presented, the next subsection gives arguments for the occurrence of the special values $-1/2$ and $3/2$ in figures \ref{fmin} and \ref{fmax}. 

\subsection{Significance of $-\frac12$ and $\frac32$}

Suppose Poisson's ratio is the same for two different pairs of directions: 
$\nu ({\bf n}, {\bf m}) = \nu ({\bf n}^*, {\bf m}^*)$. 
The pairs $({\bf n}, {\bf m})$ and $({\bf n}^*, {\bf m}^*)$ must satisfy, using \rf{e13}, 
\beq{y3}
\frac{ D({\bf n}, {\bf m}) -  D({\bf n}^*, {\bf m}^*) }{ F({\bf n})-  F({\bf n}^*)}= 2\nu \, .  
\eeq

For instance, let ${\bf n}^* = 110$, ${\bf m}^* = 001$, so that $\nu = \nu_{001}$.  Equation \rf{y3} implies that the same Poisson's ratio is achieved for directions $({\bf n}, {\bf m})$ satisfying 
\beq{y4}
  \frac{ D({\bf n}, {\bf m})  }{ 2F({\bf n})- 1/2} = \nu_{001}\, .  
\eeq
Note that this is independent of $\nu_0$ and $\chi_0$. 
We choose $\nu_{001}$ specifically because it  has been  viewed as the candidate for largest Poisson's ratio, until Ting \&  Chen (2005).   If it is {\bf not} the largest, then there must be pairs $({\bf n}, {\bf m}) $ other than $(110, \, 001)$ for which \rf{y4} holds.  However, it may be shown  using results from  \S\ref{sec2}
that the minimum of the left member in \rf{y4} is $3/2$, and 
the minimum occurs at ${\bf n} = 110$, as one might expect. 
This indicates that $\nu_{001}$ must exceed $3/2$ in order for the largest Poisson's ratio to occur for $\bf n$ other than the face diagonal $110$. 

Returning to \rf{y3},  let $\nu = \nu_{1\bar{1}0}$, then $\nu ({\bf n}, {\bf m}) =\nu_{1\bar{1}0}$ if 
\beq{y6}
  \frac{ \frac12 - D({\bf n}, {\bf m})  }{ 2F({\bf n})- \frac12} = - \nu_{1\bar{1}0}\, . 
\eeq
Using equation \rf{fol} and the previous result, 
it can be shown that the minimum of the left member in \rf{y6} is $1/2$,
 and the minimum is at ${\bf n} = 110$.  
Hence,    $\nu_{1\bar{1}0}$ must be less than  $-1/2$ in order for the smallest  Poisson's ratio to occur for $\bf n$ other than the face diagonal $110$. 
These  two results explain why the particular values $\nu_{001}=\frac32$ and $\nu_{1\bar{1}0}=-\frac12$ appear in tables 1 and 2 and in figures \ref{fmin} and \ref{fmax}.


\end{document}